\documentclass[aps,prl,twocolumn,superscriptaddress]{revtex4}
\usepackage{epsfig}
\begin{document}
\title{Magneto-transport properties of dilute granular ferromagnets}
\author{A. Cohen A. Frydman and R. Berkovits}
\affiliation{The Physics Department and the Minerva center, 
Bar-Ilan University, Ramat-Gan 52900, Israel}
\begin{abstract}
We present magnetoresistance (MR) measurements performed on quench condensed
granular Ni thin films which are on the verge of electric continuity. In
these systems the electric conductivity is believed to be governed by the
resistance between a very small number of grains. The films exihibit sharp
resistance jumps as a function of magnetic field. We interpret these
findings as being the result of magneto-mechanical distortions that occour
in single grains which act as bottlenecks in the dilute percolation network.
The observed features provide a unique measure of magnetostriction effects
in nano-grain structures as well as being able to shed light on some of the
properties of regular granular magnetic films.
\end{abstract}

\maketitle

Transport in granular metals, i.e. systems of metallic islands imbedded in
an insulating matrix, has been an active area of research for a number of
decades but many experimental findings are still far from being understood.
Among these are the temperature dependence of the resistance ($R\alpha
\exp [\frac{T_{0}}{T}]^{0.5})$ and the large noise associated with granular
films. The case in which the metallic islands are ferromagnetic is of
particular interest since these systems show magnetoresistance (MR)
properties which are reminiscent of the giant magnetoresistance phenomena
characteristic of magnetic multilayers.

A natural way to treat granular metals is by considering a Miller-Abrahams
resistor network in a similar manner to that of disordered systems \cite%
{ambegaokar}. Each pair of grains is represented by a resistor with
conductance proportional to the tunneling probability between the grains %
\cite{adkins}. In this representation the sample conductance is governed by
''critical resistances'' which act as bottlenecks for the conductance and
dominate the electric transport of the percolation network. Since the
density of these critical resistors is determined by the distribution of
resistances one can expect that as the distribution becomes very large the
distance between critical resistors (i.e. the correlation length of the
electric percolation network) grows until eventually it should approach the
sample size. In this limit the sample is critical and the behavior of any
bottleneck grain will have large influence on the total conductance. In this
paper we describe conductance measurements performed on granular Ni films in
which the transport is believed to be governed by a very small number of
grains. The samples exhibit unique sharp resistance jumps as a function of
magnetic field. We attribute these jumps to buckling type magnetostriction
deformation of bottleneck grains leading to abrupt changes in the sample
conductivity.

An established method to fabricate systems of granular metals in general %
\cite{strongin,granular bob,granular goldman,granular rich} and granular
ferromagnets in particular \cite{aviad1,aviad2,philmag} in a very controlled
way is quench condensation. In this method thin films are grown by
sequential evaporation on a cryogenically cold substrate under UHV
conditions while monitoring the film thickness and resistance. If the
samples are quench condensed on a non-passivated substrate such as SiO, they
begin growing in a granular manner so that the film is constructed of
separated islands. As more material is quench condensed, the average
distance between the islands decreases and the resistance drops. In these
samples there is a critical thickness $d_{C}$, below which no conductivity
can be practically measured. Once the thickness, $d$, of the sample is
larger than $d_{C},$ the sheet resistance, R, drops
exponentially with $d$ reflecting the tunneling nature of the transport in
these systems.

\begin{figure}\centering
\epsfxsize8.5cm\epsfbox{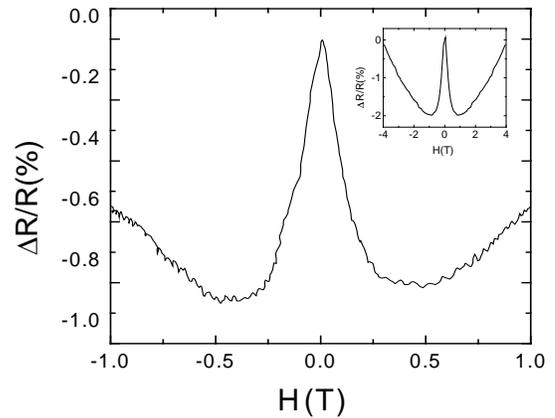} \caption
{Typical magnetoresistance for a conventional 2D quench condensed granular Ni
film having resistance of 1M$\Omega$ at T=4.2K. The insert shows a large field
MR of a similar sample showing that the positive MR trend continues to very 
large fields.
}\end{figure}

The quench condensation method provides a very sensitive control on the
sample growth process, allowing one to terminate the evaporation at any
desired stage of the material deposition and ''freeze'' the morphological
configuration. In particular, it allows one to stop the film growth at a
thickness for which a measurable conductivity first appears across the
sample ($d\approx d_{C}$). The current study was performed on quench
condensed granular Ni films which were prepared under these conditions.
Recently, we have studied a number of quench condensed granular materials
in-which deposition was terminated at the verge of electric conductivity %
\cite{future}. These show a number of features which differ from the usual
2D macroscopic systems and are characteristic of mesoscopic samples . Among
these are the following \cite{future}:

\begin{itemize}
\item Resistance switches and induced noise appear randomly (as a function
of time) in many of these samples. There are also large sample to sample
variations in the transport properties.

\item Superconducting films show magnetoresistance oscillations similar to
those which are observed in 1D superconducting granular wires \cite{herzog}.
The area associated with these oscillations is (700\AA )$^{2}$.

\item As in macroscopic granular films, these samples exhibit an
electric field dependence of the form I$\propto e^{\frac{F_{0}}{F}^{\frac{1}{%
3}}}$for large electric field, F.  However, the effective sample length
extracted from the measured I-V curves is three orders of magnitude smaller
than those of macroscopic systems.
\end{itemize}

These findings lead us to consider the possibility that despite the fact
that these samples have macroscopic dimensions ($\sim $0.5 cm$^{2}$), the
percolation network is dilute enough so that the transport is governed by a
very small area of the sample. This observation is supported by our
measurements of the magneto-transport of granular ferromagnets described
below. We name these samples ''dilute samples'' as opposed to the regular 2D
films.

The magnetoresistance of 2D quench condensed Ni has been extensively studied
in the past \cite{aviad1,aviad2,philmag}. A typical MR trace is shown in
 figure 1. The field in these experiments was applied perpendicular to the
film. The MR curve shows a resistance peak centered at H=0 followed by
negative MR up to fields of $\pm 0.5T$. This behavior is similar to that
observed in insulating granular ferromagnets prepared by co-evaporation of a
magnetic material and an insulation material \cite%
{gittelman,gerber,yang,honda,sankar} and is a result of spin dependent
tunneling between grains which have randomly oriented magnetic moments \cite%
{abeles}. Applying a magnetic field aligns these moments causing a
resistance decrease. The magnitude of the MR in our films is always of the
order of 1-2\% at T=4K. A unique feature that appears in all of the
high-resistance quench condensed films is the increase of the resistance at
high fields (H%
\mbox{$>$}%
0.5T) rather than a saturation that is expected for large H where all the
grain moments are assumed to be aligned. This finding has been a puzzling
 issue for several years. As we note below, the results presented in this
current work may shed some light on this high-field positive MR trend.

\begin{figure}\centering
\epsfxsize8.5cm\epsfbox{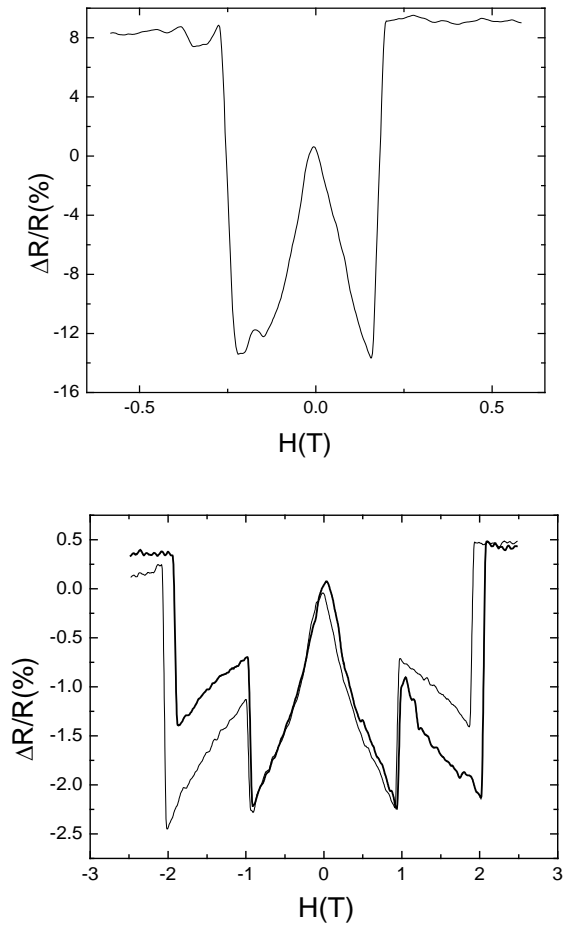} \caption
{Magnetoresistance traces of dilute granular Ni samples in
which the quench condensation was terminated close to the threshold for
electric continuity. The top frame is an example for a sample that exhibits
a single jump. The bottom frame shows the MR of a sample showing a
double-jump structure. The heavy line is for a field swept from left to
right and the light line is for the opposite sweep
}
\end{figure}
\begin{figure}\centering
\epsfxsize8.5cm\epsfbox{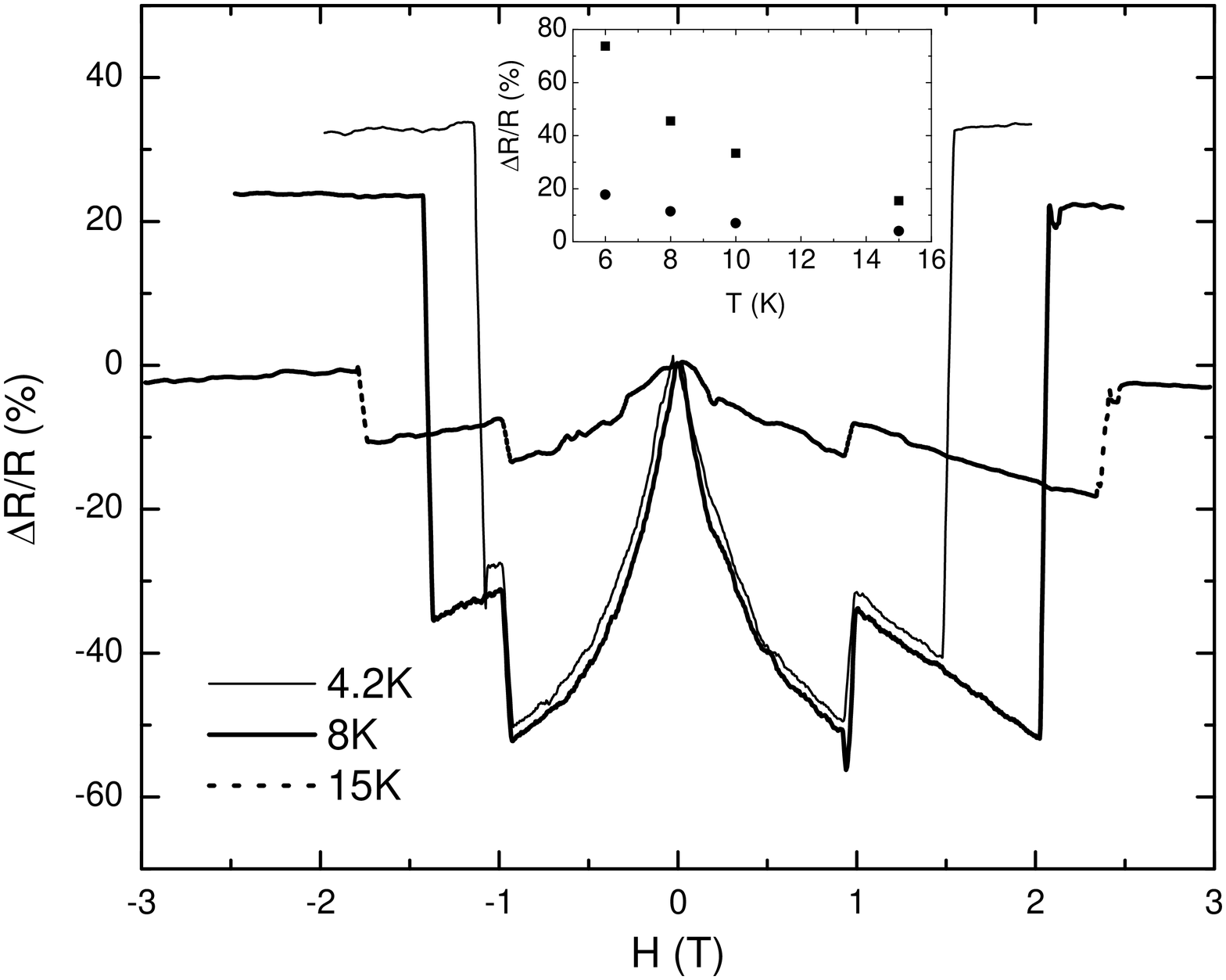} \caption
{The temperature \ dependence of a MR curve. The insert shows
the dependence of the amplitude of the two resistance jumps (circles for the
first jump and squares for the second) on temperature. The amplitude
extrapolates to zero at around T=35K.
}\end{figure}

The MR curves of the dilute samples show several unique features that are
not seen in the regular granular systems. These are depicted in figures 2
and 3 where it is seen that at low fields the samples exhibit a negative MR
similar to those of regular 2D films. However, at specific magnetic fields the
resistance switches abruptly to a higher value. In some cases we observe a
single resistance jump, such as that seen in the top frame of figure 2,
after which the resistance saturates at a high resistance value. In other
cases we observe two jumps (figures 2 bottom frame and figure 3). The first
is a relatively small resistance increase followed by the continuation of
the negative MR trend and the second is a large resistance jump leading to
saturation at a high value. It is worth noting that although the qualitative
behavior at low fields is similar to that of usual 2D granular systems, the
magnitude of the negative MR in our samples can be much larger. While the MR
reported in previous works on granular ferromagnets rarely acceded 2\%, we
commonly observe effects of the order of 10\% and in some cases the MR is as
large as 60\% (see figure 3). Clearly, in these films there is a large
sample-to-sample variation in the magnitude of the negative MR and of the
sharp jumps as well as the in the magnetic field at which the jump occurs.
This feature further reflects the mesoscopic nature of these geometries.

A number of properties of the MR curves should be noticed. First, although
the traces are very reproducible, the jumps are hysteretic as seen in figure
2. Sweeping the field in opposite directions produces mirror images of MR.
Second, raising the temperature results in two clear trends (figure 3): The
magnitude of the jumps decreases with raising the temperature in a similar
manner to that of the negative MR at low fields, and, at the same time the
magnetic field of the jumps shifts to higher values. Finally, perhaps the
most surprising finding is the fact that the saturated resistance value at
high magnetic fields is larger than that of the value at H=0. At zero field
the magnetic grain moments are believed to be totally random and the
resistance is expected to be maximal. The high resistance value at high
magnetic fields is therefore a very unexpected result.

In an attempt to understand the above behavior we observe that the electric
properties of these samples may be dominated by a small number grains which
are characterized by the largest inter-grain distance that participate in
the conduction paths. We have considered the possibility that the jumps are
a result of magnetic moment flip of such a bottleneck grain. Recently,
Meilikhov \cite{grainflip} considered the behavior of small single-domain
grains in an external magnetic field and showed that their magnetic moment
may flip abruptly if the initial angle between the magnetic field and the
easy axis of the grain is larger than $\pi /2.$If a bottleneck grain is
subject to such magnetic moment flip the MR of the entire sample conductance
would be subject to sharp features.

However, the above model seems to be inconsistence with a number of our
observations. In the first place this scenario could naturally explain sharp
resistance {\it decreases.} Our samples always exhibit sharp {\it increases}
in the resistance. Furthermore, the resistance value after the jump is
higher than that at H=0 where the grain moment orientations are assumed to
be completely random. Secondly, a pre-requirement for a resistance jump is
that the bottleneck grain moment orientation have a larger-than-$\pi /2$%
-angle with the magnetic field direction. Our grains are believed to be in
the superparamagnetic state, and it is hard to see how this condition is
fulfilled. But even if such a situation occurs, computer simulations on such
a granular configuration show that if the grain moment begins with such an
angle, the initial MR would be positive rather than negative. In addition to
the above, according to the Meilikhov picture the field at which the jump
occurs should decrease with increasing temperature since the thermal energy
assists in overcoming the anisotropy energy. This is opposite to the trend
observed in figure 3.

\begin{figure}\centering
\epsfxsize8.5cm\epsfbox{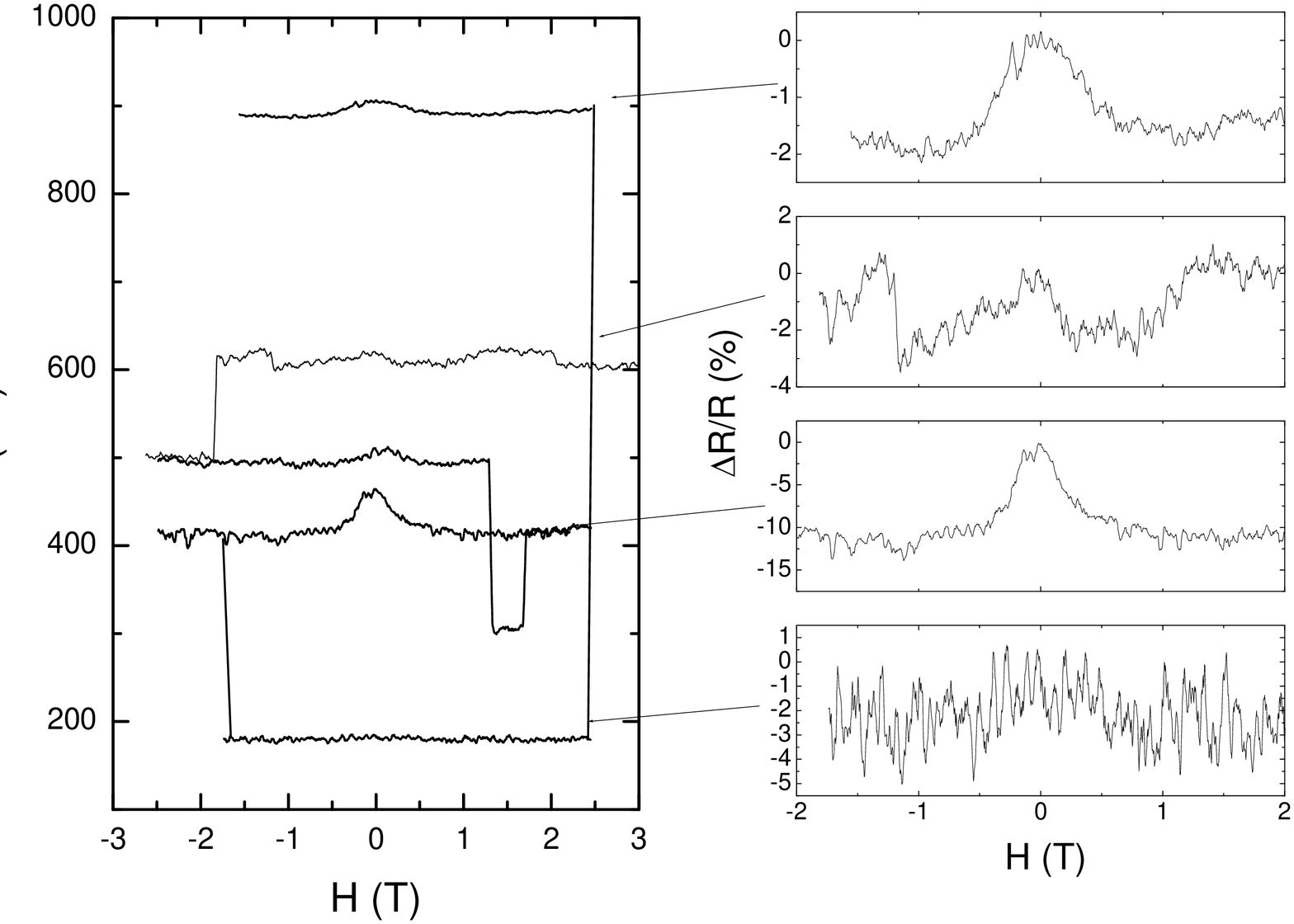} \caption
{Resistance versus magnetic field of a dilute sample showing
pronounced switching behavior while sweeping the magnetic field back and
forth many times. The right frames are the zoom on the MR curves of the
different branches
}\end{figure} 

We have also considered the possibility that the resistance jumps are
associated with switches between different current trajectories leading to
sudden resistance changes. Such a process also doesn't seem to explain our
observations because of the reproducibility of the data. The jumps are
always upwards, they are always of the order of magnitude of the negative MR
itself and the curves trace each-other when sweeping the field many times.
Random switches between different trajectories are expected to cause either
increase or decrease of the resistance with the same probability, to be
random in size and magnetic field and to be non-reproducible. Indeed, a
number of our samples showed much larger resistance switches which were
random both in field and in sign. An example is shown in figure 4. It is
seen that after each switch the magnetoresistance curve follows a different
trace having different MR magnitudes (ranging between 0.5-13 \%). In
particular, one of the traces shows a resistance jump similar to the effects
described above. It seams reasonable to assume that this sample (unlike that
of figures 1,2 and 3) is characterized by a number of bottleneck trajectories
and the current switches randomly between these paths leading to the
different MR fingerprints.

The above considerations lead us to suggest that the effects seen in our
samples are due to magneto-mechanical deformation of the grains.
Magnetostriction (MS) in bulk Ni is well known and widely studied but not a
lot is known about such effects in nano-structures. Geometry changes in
small grains are extremely difficult to measure and, in granular films which
are achieved by co-evaporation, the grains are imbedded in an insulating
matrix and any MS effects depend strongly on the elasticity of this
insulator. The quench condensed systems provide us with a unique opportunity
to study MS effects on small grains since they are free standing particles.
We envision that as the grains start aligning parallel to each other, all
perpendicular to the substrate, and dipole-dipole interactions cause
magnetic repulsion between the grains. At some magnetic field, a buckling
instability may occur and a small MS distortion of the grains will take
place causing a conductance decrease. Since the conductance is achieved by
tunneling, any small distortion (even by fractions of an \AA ) in the
geometry of the grains that belong to a conduction bottleneck will have a
large effect on the conductivity. We estimate the magnetostatic interactions
between two neigbouring parallelly aligned grains to be approximately 70K
(about an order of magnitude larger than the thermal energy of the system at
the relevant measurements). Our experience shows that increasing the
temperature of quench condensed grains above 40K causes annealing and
irreversible morphology changes in the sample. It seems reasonable,
therefore, that magnetic forces in our samples could also have a
considerable effect on the morphology of the grains. According to this
model, if an external field aligning the grains is applied perpendicular to
the substrate, two opposing effect influence the resistance. One is the fact
that the tunneling probability is higher for aligned grains. This acts to
lower the resistance causing a negative MR background. The second is a
magnetostatic repulsion causing the inter-grain distance to increase
slightly, thus decreasing the tunneling probability at the buckling
instability fields.

The model described above can shed light on some of the features observed in
the MR of usual 2D quench condensed granular samples. As noted above, one of
the puzzling features in these samples is the increase of the MR at high
magnetic fields at which the resistance is expected to have reached
saturation (see figure 1). This anomaly is observed mainly when the
resistance of the granular films is relatively high. Such behavior can be
explained by magnetic repulsion between grains causing minute geometrical
changes which decrease the inter-grain tunneling probabilities causing
resistance increases. In the dilute systems, this manifests itself as sharp
resistance jumps while in regular samples one observes a continuous mild
increase in resistance with increasing field. Such an effect can be
particularly expected in quench condensed samples in which the grains are
free standing on a substrate separated by vacuum. In granular samples, which
are achieved by co-evaporation of metal and insulating materials,
geometrical modifications are less probable.

In summary, we have fabricated and measured dilute granular Ni films in
which the transport is believed to be governed by a very small number of
grains. These samples exhibit sharp resistance increases at discrete
magnetic fields. In order to explain these findings we proposed a
qualitative model which involves buckling of bottleneck grains due to
magnetostatic repulsion. The unique geometry of our samples enables one to
detect the properties of a single grain even if the sample has macroscopic
dimensions. Our granular samples also provide a unique opportunity to study
magneto-mechanical effects on nanometer sized magnetic structures. The
results obtained on these extreme samples shed some light on the behavior of
the ferromagnetic granular samples in general and they may also be relevant
for the development of new types of magneto-machanically based single-grain
devices.

We gratefully acknowledge illuminating discussions with R.C. Dynes, Z.
Ovadyahu and M. Pollak. This research was supported by the Binational
USA-Israel fund grant number and by the Israeli academy of science.

\end{document}